# A hybrid vanadium fluoride with structurally isolated $S = 1$ kagome layers

Farida H. Aidoudi[1], Lewis J. Downie[1], Russell E. Morris[1], Mark de Vries[2] and Philip Lightfoot[1]*

1. EaStChem, School of Chemistry, University of St Andrews, St Andrews, Fife KY16 9ST, UK.
2. EaStChem, School of Chemistry, University of Edinburgh, EH9 3JZ, UK.

E-mail: pl@st-and.ac.uk

**A new organically-templated vanadium (III) fluoride, $(NH_4)_2(C_2H_8N)[V_3F_{12}]$, has been prepared using an ionothermal approach. This compound has a unique layered structure featuring distorted $S = 1$ kagome planes separated by the cationic species. The compound exhibits magnetic frustration, with a canted antiferromagnetic ground state. On further cooling within the ground state a pronounced change in magnetisation kinetics is observed.**

The kagome lattice antiferromagnet is perhaps the most famous example of a geometrically-frustrated magnetic system[1,2]. The ideal kagome antiferromagnet consists of layers of corner-sharing equilateral triangles, with antiferromagnetically-coupled magnetic cations at each node. The inherent magnetic frustration in this system can be seen when one attempts to place 'spins' on each node of the triangles such that all near-neighbour interactions satisfy antiparallel (ie. antiferromagnetic, AFM) alignment. Not all preferred interactions can be satisfied simultaneously, leading to a vastly-increased number of compromise spin configurations of equally low energy. This frustration can result in various states that lie beyond the standard notions of long-range magnetic ordering (LRO) and spontaneous symmetry breaking, leading to either spin glass, spin ice and spin liquid states[3,4]. The inhibition of LRO is particularly acute in systems with 'quantum spins' (ie. $J = S = ½$ for the magnetic cations) for which quantum fluctuations combine with frustration to suppress LRO down to the lowest-accessible temperatures, in favour of quantum spin liquid ground states (QSLs).

There is now growing evidence that these long-sought QSL states exist in recently studied $S = ½$ kagome antiferromagnets including herbertsmithite[5,6] and kapellasite[7] (both polymorphs of the composition $ZnCu_3(OH)_6Cl_2$) and the vanadium oxyfluoride $[NH_4]_2[C_7H_{14}N][V_7O_6F_{18}]$ (DQ-VOF)[8,9]. Analogous systems with $S = 1$ spins usually freeze into LRO states[10-16] (full details are given in ESI, Table S1) because here quantum fluctuations play a diminished role and probably more importantly, due to single ion anisotropy effects, which start to play a role for $S > ½$. Nevertheless, the diverse range of behaviour exhibited by the few examples so far, and the various theoretically proposed states for the related $S = 1$ triangular lattices[17,18], demonstrates that further, chemically different, examples of $S = 1$ kagome magnets are of significant interest.

In this paper we report a novel example of this family, in the compound $(NH_4)_2(C_2H_8N)[V_3F_{12}]$ **1**. This study forms part of our ongoing efforts to explore and understand the complex solvothermal/ionothermal chemistry of vanadium fluoride-based systems[8,19-21]. Specifically, compound **1** was isolated by using identical chemistry[¶] to that in the preparation of the QSL compound DQ-VOF, except that the temperature of the reaction was increased from 170 to 210 °C and the time was increased from 1 day to 5 days. Remarkably, this adjustment in reaction conditions produces a second example of a vanadium fluoride-based kagome system, but with major differences in both structure and electronic behaviour when compared to DQ-VOF. The subtlety and richness of these vanadium fluoride systems is further emphasised as we also report the occurrence of two side-products formed in the preparation of **1**, both of which have kagome-related structural features.

The crystal structure[§] of **1** is shown in Figure 1. Near-ideal kagome layers of composition $[V_3F_{12}]_n^{3n-}$ exist in the *bc* plane, and these are separated by the ammonium and ethylammonium counterions. The kagome layer consists of three crystallographically-distinct $VF_6$ octahedra, sharing in-plane corners. As discussed in our previous studies[8,20], increasing the reaction temperature in these systems favours reduction of vanadium to the trivalent state; bond-valence sum (BVS) analysis (see ESI) and magnetic data[†] show that in this case all three distinct V sites correspond to $V^{3+}$, rather than the more oxidised state ($6V^{4+}/1V^{3+}$) occurring in DQ-VOF[8]. In each of the three octahedra, the geometry is tetragonally shortened, such that the apical V-F distances are in the range 1.86 – 1.89 Å and the in-plane distances are 1.94 – 1.99 Å; the underbonding at the apical sites is partially alleviated by acceptance of a network of hydrogen bonds from the interlayer cations. Considering the vanadium sites only, the kagome net is almost flat, with a slight buckling at V3 (V1-V3-V2 = 173.16(2)°) and also quite regular, with the three unique V---V edges being 3.711, 3.722 and 3.739 Å. However, the octahedral units are considerably 'tilted' around the *bc* plane, with the tilt angles in the range 13.1 – 21.5°.

Magnetic data for **1** up to 300 K show Curie-Weiss paramagnetism above 20 K (Figure 2). Taking into account a small temperature-independent contribution to the susceptibility, a value of $\mu_{eff}$ = 2.45 $\mu_B$ was found for the effective moment. For $S$ = 1 this implies g = 1.73, a considerable reduction of the spin-only value of 2 which is not uncommon for V moments. The negative Weiss temperature, $\theta$ = -30 K, implies predominantly AFM interactions within the kagome layers. The zero-field cooled magnetic susceptibility data (Figure 3) point to a freezing transition below 15 K with a small ferromagnetic moment of only 10% of the saturated moment of $S$ = 1 spins with $g$ = 1.73. When such a small ferromagnetic moment arises in an antiferromagnetic system this points to an antiferromagnetic ground state, most likely with some kind of 120 degree compromise structure with a small out-of-plane component (phase 2). Within the constrained subset of available spin configurations there is apparently a freedom for the out-of-plane spin component to align with an applied field, giving rise to a small ferromagnetic moment. A hysteresis loop at 5 K (Figure 4) shows that this moment is remanent, with a coercive field of ~ 800 Oe. The diversion between the field-cooled and zero-field cooled susceptibility below 5 K (Figure 3) points to a second transition to what is likely the ground state (phase 1). This phase is characterised by a vastly-increased coercive field - well in excess of 10 kOe as is obvious from the 2 K hysteresis loop in Figure 4, while the field-cooled data show there is no change to the maximum remanent moment on cooling from phase 2 to phase 1. This suggests that kinetic barriers between spin configurations in the ground state manifold increase as thermal fluctuations are progressively suppressed.

Comparing compound **1** to the previously known examples of purely $V^{3+}$ materials with structurally isolated kagome layers. ie. the jarosites, $MV_3(SO_4)_2(OH)_6$ (ESI Table 1), it is interesting to note the contrasting intra-layer magnetic interactions: FM in the jarosites AFM in **1**. This may be rationalised by comparing the local geometries of the $V^{3+}$ octahedra. In the jarosites, the 'capping' of the kagome layer by the sulphate group necessitates *longer* apical bonds within the $VO_6$ octahedra[10,11]; ie. a tetragonal *elongation*. In **1**, the opposite is observed – a tetragonal *contraction*, due to the lack of additional covalent coordination at the apical site. This means that the d-orbital splittings will differ and the preferred configurations will be $(d_{xz})^1(d_{yz})^1$ for the jarosites and $(d_{xy})^1(d_{xz},d_{yz})^1$ for **1**. Nocera and co-workers[22] suggest that the superexchange in the $V^{3+}$ jarosites (in contrast to $d^3$ or $d^5$ analogues) is dominated by $d_{xz}(\pi)$-p(O)-$d_{xz}(\pi)$ pathways, leading to net FM interactions. In **1**, the $d_{xy}(\sigma)$-p(F)-$d_{xy}(\sigma)$ pathway will also contribute, leading to the observed AFM interaction becoming favoured. This restricted orbital occupancy will also give rise to a single-ion anisotropy, restricting the spins into the equatorial plane of the $VF_6$ octahedra. Because these equatorial planes are buckled out of the kagome planes a net out of plane magnetisation can arise in a nominally staggered ground state.

Although **1** can be prepared in an almost pure state (see ESI for details), several of our reactions revealed small amounts (< 5%) of one of two secondary phases. Both of these adopt structures comprising well-separated $[VF_4]_n^{n-}$ layers. A well-determined crystal structure of $NH_4VF_4$ (**2**) (ESI) reveals a layer of tetragonal tungsten bronze (TTB) type, rather than the hexagonal tungsten bronze (HTB) type (equivalent to the kagome layer). This phase is isostructural to the low-temperature polymorph of $NH_4AlF_4$[23]. The second impurity phase provided only a partial crystal structure determination. Nevertheless, the gross features of the structure (ESI) reveal a kagome layer similar to that in **1**, but this time template by interlayer $NH_4^+$ and methylammonium cations. In both cases, it can be seen that the higher reaction temperature (compared to that used in the preparation of DQ-VOF) has led to reduction to $V^{3+}$, and also to differing degrees of breakdown of the organic species. These subtle differences in reaction outcomes suggest further avenues for exploration and control of the solvothermal chemistry of vanadium (oxy)fluoride systems.

**Acknowledgements.** We thank Dr Catherine Renouf, Dr Laura McCormick and Mr Samuel Morris for assistance in collecting the diffraction data at ALS. FHA acknowledges support from the EPSRC Doctoral Prize Fellowships scheme (EP/J500549/1). REM is a Royal Society Wolfson Merit Award holder. The ALS is supported by the Office of Basic Energy Sciences, U.S. Department of Energy, under Contract No. DE-AC02-05CH11231.

**Notes and References.**

¶ Synthesis: 0.124 g of $VOF_3$ (1 × $10^{-3}$ mol, Sigma Aldrich) was weighed into a 30 mL Teflon-lined cup and 0.1 mL of HF(48 wt% in $H_2O$) (2.76 mmol, Sigma Aldrich), was added. To this, 4g (~$10^{-2}$ mol) of the ionic liquid (EMIM)($Tf_2N$) was added along with 0.111 g of quinuclidine (1 × $10^{-3}$ mol, Sigma Aldrich), and the sealed vessel heated at 210°C for 5 days. Upon cooling to room temperature, the brown crystals were filtered, washed with methanol and dried in air for 24 hrs. The ionic liquid 1-ethyl-3-methylimidazolium bis(trifluoromethylsulfonyl)imide, (EMIM)($Tf_2N$) was prepared by standard methods (ESI). Elemental analysis (Calc: %C: 5.18, %H: 3.45, %N: 9.07; Found: %C: 5.28, %H: 3.36, %N: 8.92). Phase purity was also confirmed by comparison of observed and simulated powder X-ray diffraction patterns.

§ Crystal data for [NH$_4$]$_2$[C$_2$H$_8$N][V$_3$F$_{12}$]: $M_r$ = 463, monoclinic, space group P2$_1$/m, Z = 2, $\rho_{calcd}$ = 2.322 g cm$^{-3}$, $F(000)$ = 452, $\theta_{max}$ = 69.3°, $a$ = 6.932(5) Å, $b$ = 12.876(8) Å, $c$ = 7.444(5) Å, , $\beta$ = 94.550(7)°, $V$= 662.3(8)Å$^3$, T = 100 K, $\lambda$ = 0.77490 Å (synchrotron). A total of 9941 reflections were collected, of which 2218 were independent ($R_{int}$ = 0.28). $R_1$ = 0.0756, w$R_2$ = 0.1911, GOF = 1.000 for 115 parameters. CCDC 973935-36. Powder XRD was carried out on a Stoe STADI/P diffractometer using Cu K$_{\alpha1}$ X-rays. Single crystal X-ray diffraction data were collected at station 11.3.1 of the Advanced Light Source (ALS) at Lawrence Berkeley National Laboratories, California using a Bruker APEX II CCD diffractometer. The structure was solved by direct methods and refined by full-matrix least-squares techniques, using the SHELXS, SHELXL and WinGX packages.

† Magnetic susceptibility data for **1** were collected on a Quantum Design MPMS SQUID. Data were recorded in a 2000 Oe field while warming the sample from 2 to 300 K in sweep mode, following consecutive zero field cooling (ZFC) and field cooling (FC) cycles. Data were normalized to the molar quantity of the sample, and corrected for any diamagnetic contributions. Temperature-dependent magnetisation data were collected at various magnetic fields ranging from 100 Oe to 10 KOe, field- dependent magnetisation data were collected at three different temperatures (2K, 5K and 30K).

**Figures**

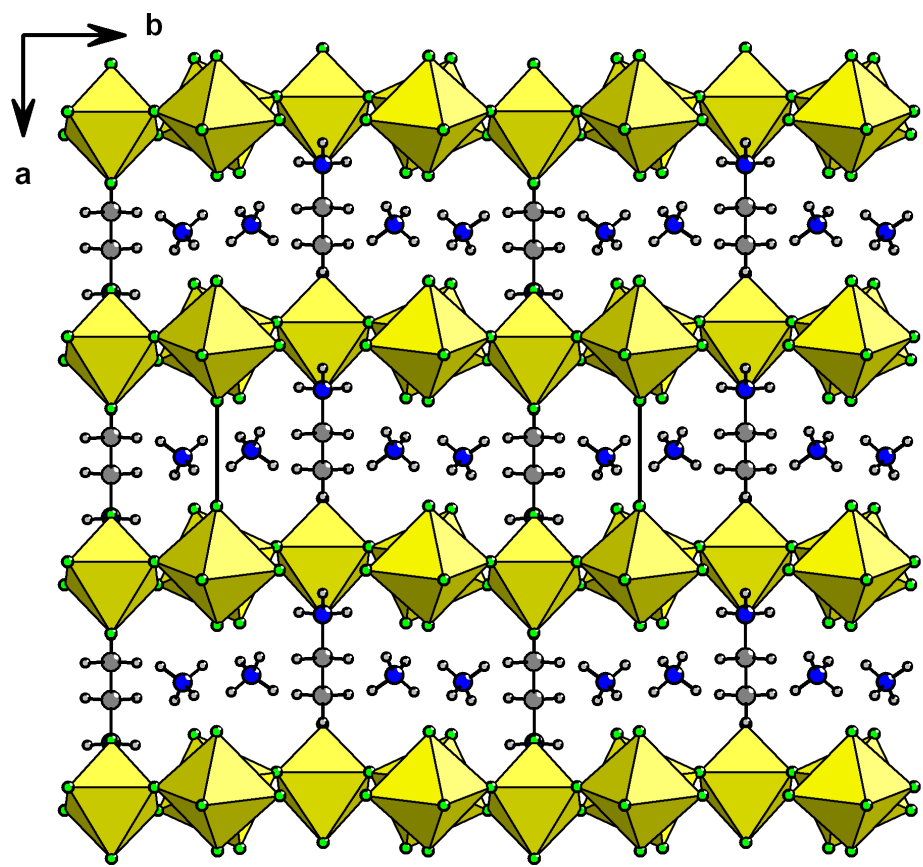

**Figure 1.** Crystal structure of **1**, showing the separated kagome planes and interlayer cationic species.

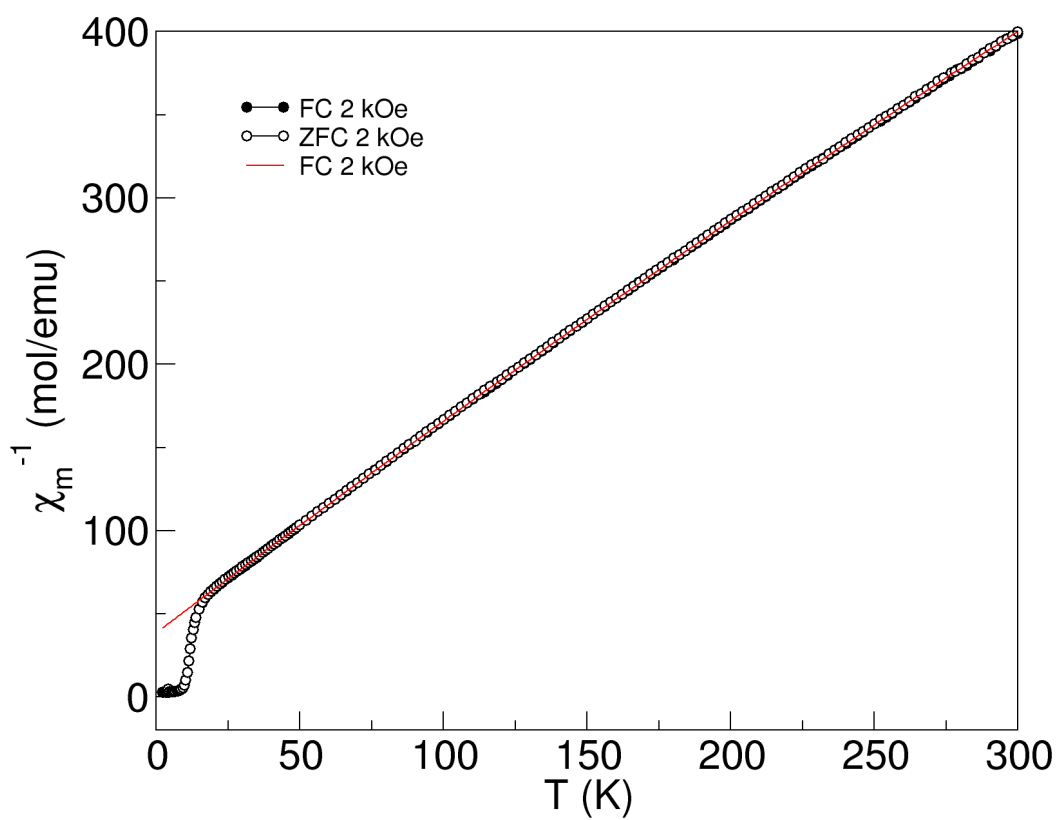

**Figure 2.** Inverse molar susceptibility (1/χ) versus T for **1**, showing Curie-Weiss fit above 20 K. Fitted to the expression $\chi_m = \chi_0 + C/(T-\theta)$, with $\chi_0 = 2.2 \times 10^{-4}$ emu mol$^{-1}$ V$^{3+}$, $C = 0.75(1)$ emu K mol$^{-1}$, $\theta = -30(5)$ K.

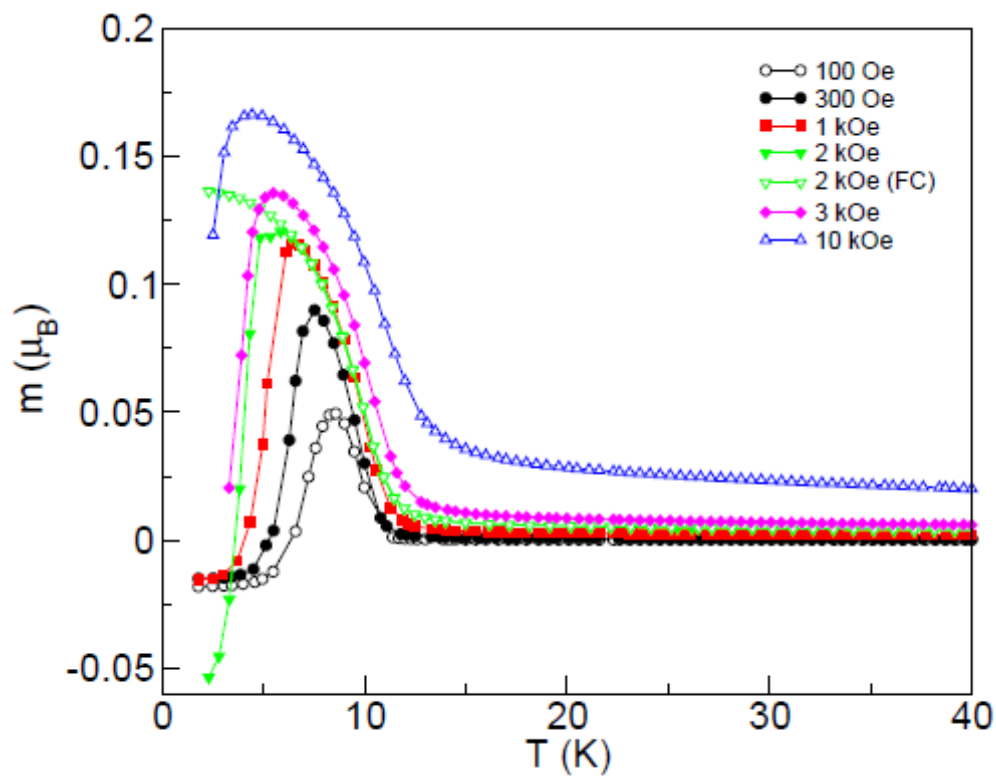

**Figure 3.** Zero-field cooled (and field-cooled at 2 kOe) magnetisation versus T.

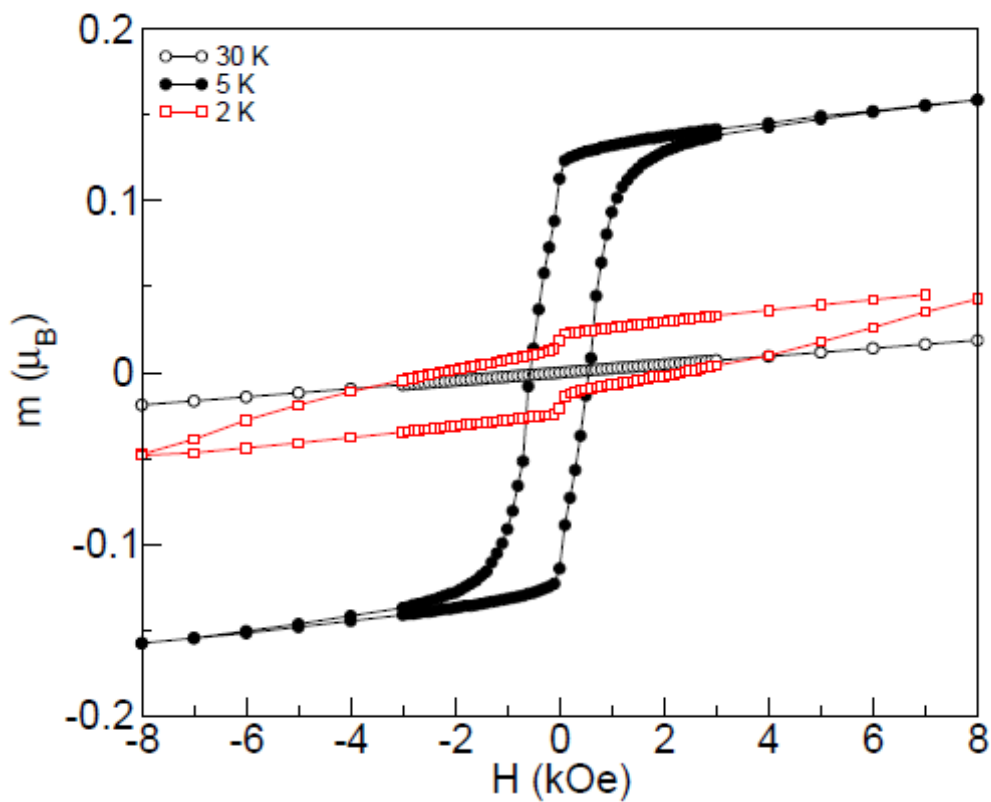

**Figure 4.** Magnetisation versus field, showing differing hysteresis behaviour at 5 K and 2 K.

# SUPPLEMENTARY MATERIAL

**Synthesis of 1-ethyl-3-methylimidazolium bis(trifluoromethylsulfonyl)imide (EMIM Tf$_2$N, mp = -3°C):**

EMIM Tf$_2$N was synthesised by an anion exchange between the IL 1-ethyl-3-methylimidazolium bromide "EMIM Br" and lithium bis(trifluoromethylsulfonyl)imide according to the literature procedure.[1]

**Table S1.** Known examples of $S = 1$ kagome systems

| Compound | Intralayer octahedral connectivity | Interlayer connectivity | Layer-layer distance (Å) | Curie-Weiss Temp. $\theta$ (K) / Freeze temp (K) | Magnetic properties | Ref |
|---|---|---|---|---|---|---|
| NaV$_3$(SO$_4$)$_2$(OH)$_6$ | V-(OH)-V vertices | Indirect: SO$_4$ caps, Na separators | 5.6 | +53 / 33 | Strong FM intra-layer, overall AFM | 10 |
| MV$_3$(SO$_4$)$_2$(OH)$_6$ (M = K, Rb, NH$_4$, Tl) | V-(OH)-V vertices | Indirect: SO$_4$ caps, M separators | 5.8-6.0 | +53 / ~ 30 | Strong FM intra-layer, overall AFM | 11 |
| YCa$_3$(VO)$_3$(BO$_3$)$_4$ | V-O-B-O-V indirect | Direct V-O | 2.9 | -453/ spin liquid (?) | Frustrated; strong AFM inter-layer, weak FM intra-layer | 12 |
| KV$_3$Ge$_2$O$_9$ | V-O-V edges | Direct: Ge-O double layer | 6.9 | -250 / 70 | Frustrated, $S = 1$, $m_S = 0$ ground state. | 13 |
| NaV$_6$O$_{11}$# | V-O-V edges | Direct: V-O double layer | 6.5 | -81 +64 / 50 - 60 | Metallic, FM | 14 |
| [C$_6$H$_8$N$_2$][NH$_4$]$_2$[Ni$_3$F$_6$(SO$_4$)$_2$] | Ni-F-Ni vertices | Indirect: SO$_4$ caps, molecular separators | 8.7 | -60 / 10 | Frustrated; canted AFM | 15 |
| BaNi$_3$(OH)$_2$(VO$_4$)$_2$ | Ni-(O/OH)-Ni edges | Indirect: VO$_4$ caps, Ba separators | 7.2 | +10/19 | Spin glass | 16 |
| (NH$_4$)$_2$(C$_2$H$_8$N)[V$_3$F$_{12}$] | V-F-V | Indirect: molecular separators | 6.9 | -30/15 | Canted AFM | This work |

# Mixed valence compound containing only V$^{3+}$ in the kagome layers

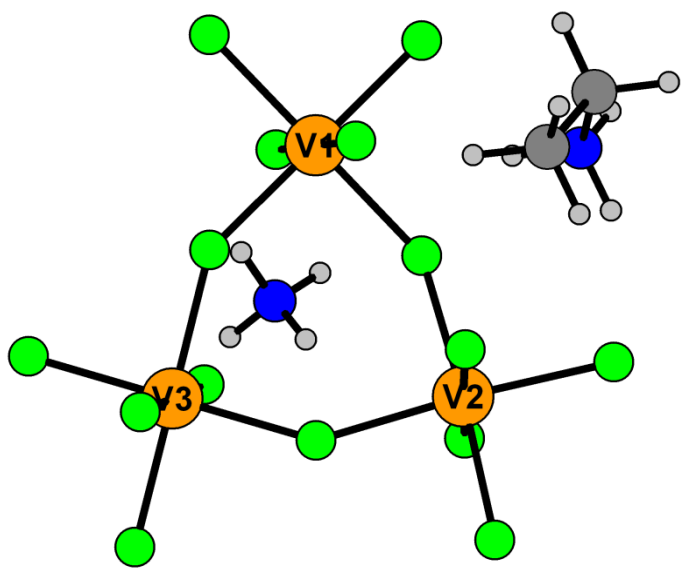

**Figure S1.** Building unit found in **1**.

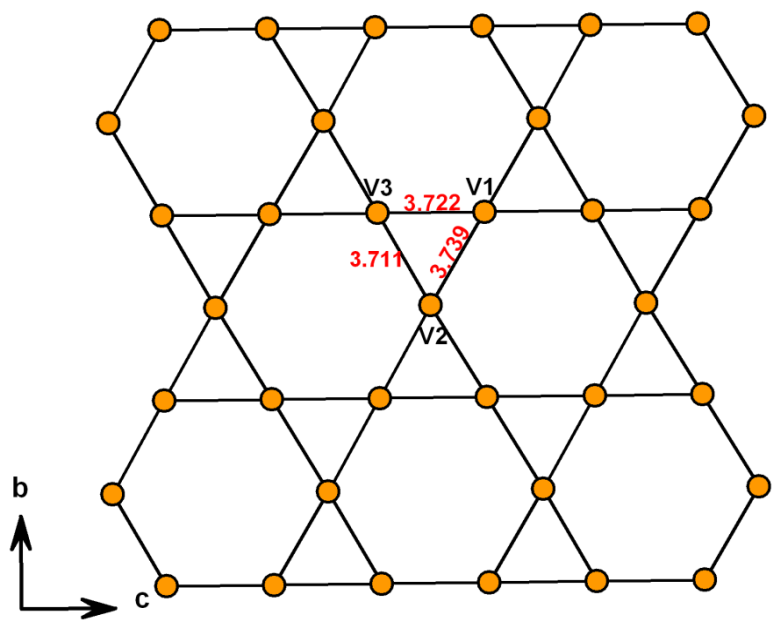

**Figure S2.** The kagome lattice found in **1**.

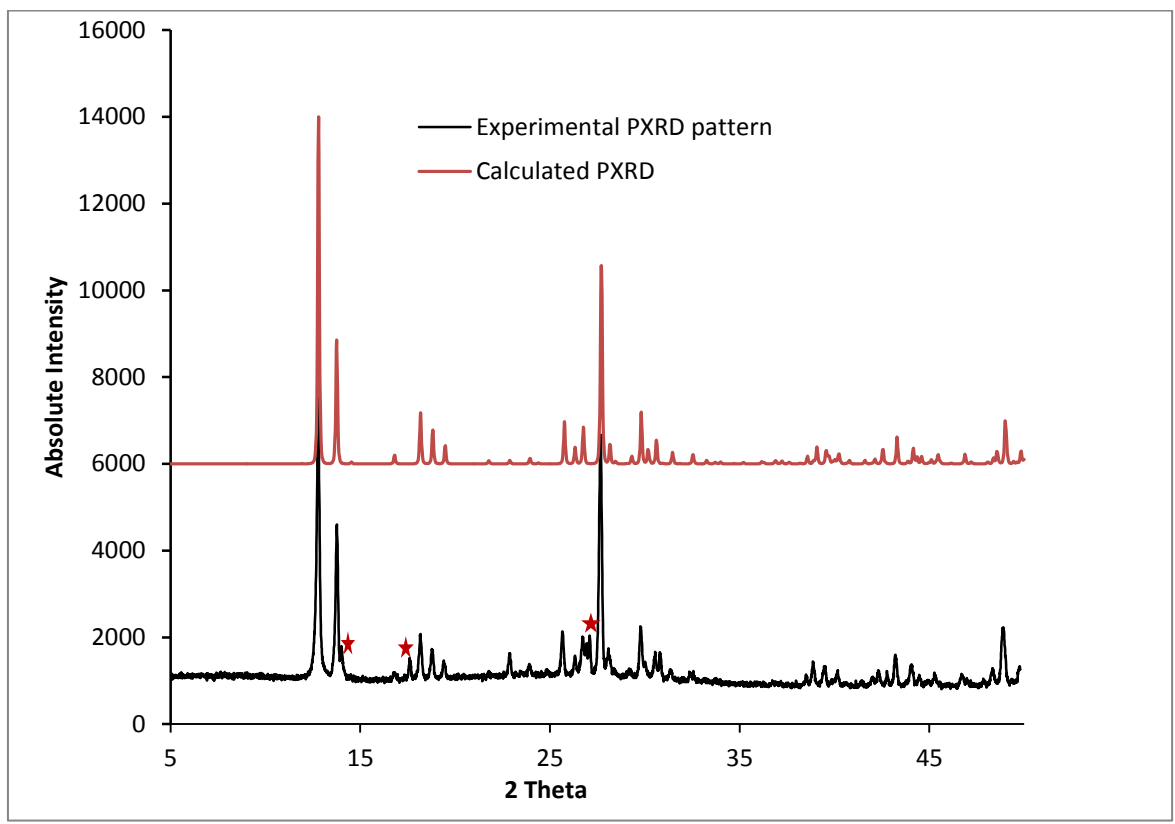

**Figure S3.** Calculated (above) and experimental (below) PXRDs for **1**. (red stars show small impurity of **2**.

Bond valence sums[2] were calculated using the program Valist.[3]

**Table S2.** Selected bond lengths (Å) and Bond Valence Sums $S_i$ for **1**

| Bond | Bond length (Å) | $S_{ij}$ | Bond | Bond length (Å) | $S_{ij}$ | Bond | Bond length (Å) | $S_{ij}$ |
|---|---|---|---|---|---|---|---|---|
| V1—F1$^i$ | 1.941(2) | 0.524 | V2—F3 | 1.9714(19) | 0.483 | V3—F1 | 1.987(2) | 0.463 |
| V1—F1 | 1.941(2) | 0.524 | V2—F3$^{ii}$ | 1.9714(19) | 0.483 | V3—F1$^{iii}$ | 1.987(2) | 0.463 |
| V1—F2 | 1.892(2) | 0.598 | V2—F4 | 1.888(3) | 0.605 | V3—F5$^{iii}$ | 1.950(2) | 0.512 |
| V1—F2$^i$ | 1.892(2) | 0.598 | V2—F5 | 1.964(2) | 0.493 | V3—F5 | 1.950(2) | 0.512 |
| V1—F3 | 1.979(2) | 0.473 | V2—F5$^{ii}$ | 1.964(2) | 0.493 | V3—F7$^{iii}$ | 1.870(2) | 0.635 |
| V1—F3$^i$ | 1.979(2) | 0.473 | V2—F6 | 1.864(3) | 0.645 | V3—F7 | 1.870(2) | 0.635 |
| | ∑ **V1** = | 3.19 | | ∑V2 = | 3.20 | | ∑V3 = | 3.22 |

(i) -x, -y, 1-z; (ii) x, 0.5-y, z; (iii) -x, -y, -z.

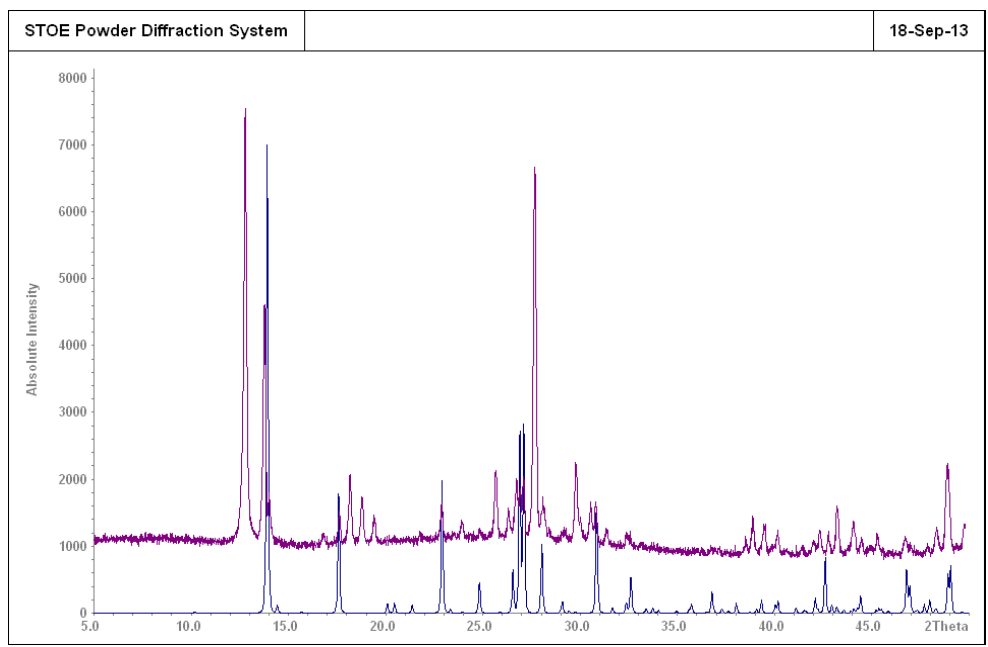

**Figure S4.** Calculated PXRD (below) for **2** and experimental PXRD (above) for **1** (Calculated PXRD of **2** fits well to the small impurity phase found in the experimental PXRD of **1**)

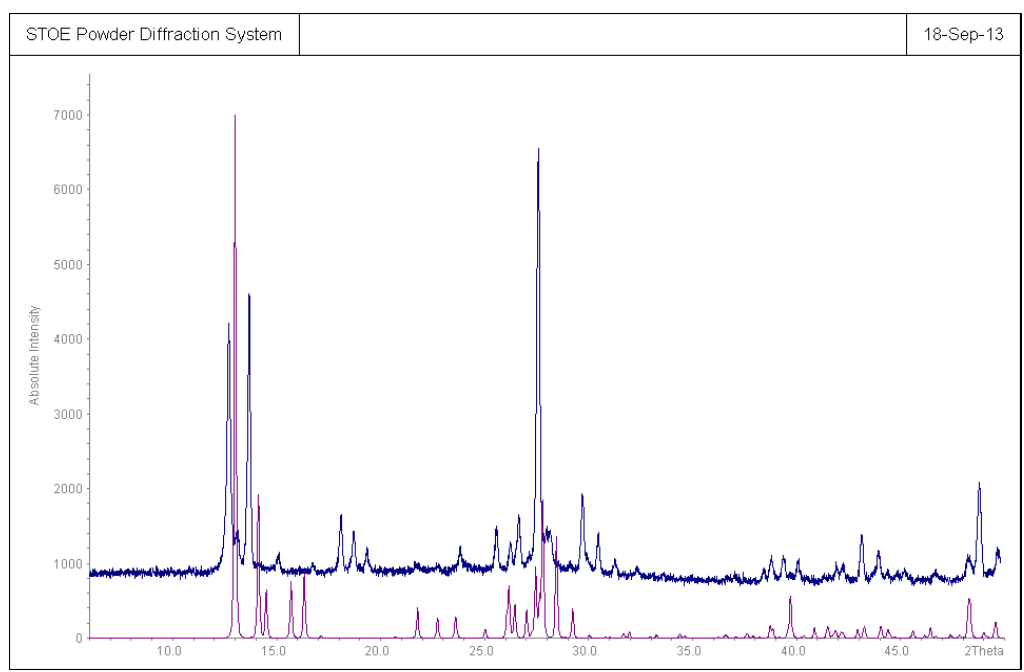

**Figure S5.** Calculated (below) PXRD for **3** and experimental (above) PXRD for **1**. (Calculated PXRD of **3** fits well to the small impurity phase found in the experimental PXRD of **1**)

Note that the experimental PXRDs for **1** in Figure S4 and Figure S5 are for samples obtained from two different reactions. Although the two reactions were carried out using the same conditions, very small amount (< 5%) of different impurity phase was present in each case.

**Table S3** Crystallographic Data for **2** and **3**.

| | 2 | 3 |
|---|---|---|
| Formula | [NH$_4$][VF$_4$] | [NH$_4$][CH$_6$N]$_2$[V$_3$F$_{12}$] |
| Fw/g/mol | 144.98 | 463 |
| Space group | $P4_2/ncm$ (4) | $P\bar{1}$ |
| $a$ /Å | 12.270(4) | 7.5410(10) |
| $b$ /Å | 12.270(4) | 7.5380(10) |
| $c$ /Å | 12.723(4) | 14.1890(10) |
| $\alpha$ /° | 90 | 74.590(10) |
| $\beta$ /° | 90 | 87.180(10) |
| $\gamma$ /° | 90 | 59.940(10) |
| $V$ / Å$^3$ | 1915.5(14) | 669.57(13) |
| Z | 20 | 2 |
| Crystal size /mm | 0.03×0.03×0.03 | 0.05×0.04×0.01 |
| Crystal shape and colour | Brown prism | Brown Plate |
| F(000) | 1400 | |
| R$_{int}$ | 0.0422 | |
| Obsd data [$I>2\sigma(I)$] | 931 | |
| Data/restraints/parameters | 937/0/103 | |
| GOOF on F$^2$ | 1.195 | |
| R1, $w$R2 (I >2 $\sigma$(I)) | 0.0194, 0.0508 | |
| R1, $w$R2 (all data) | 0.0195, 0.0509 | |
| Largest diff. peak / hole | 0.346/-0.519 | |

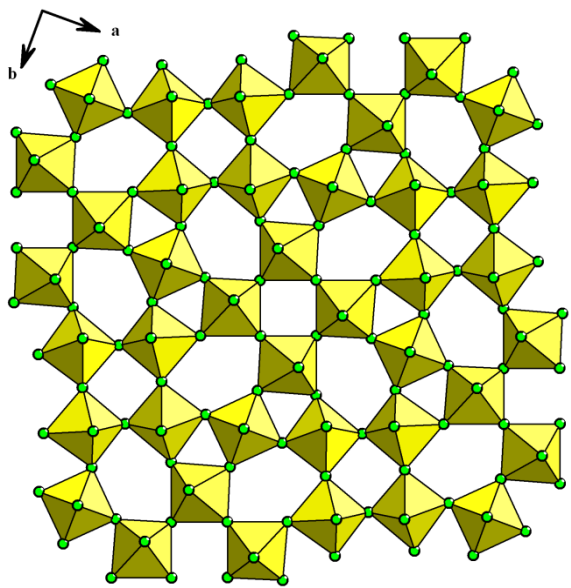

**Figure S6.** TTB layer type found in **2**.

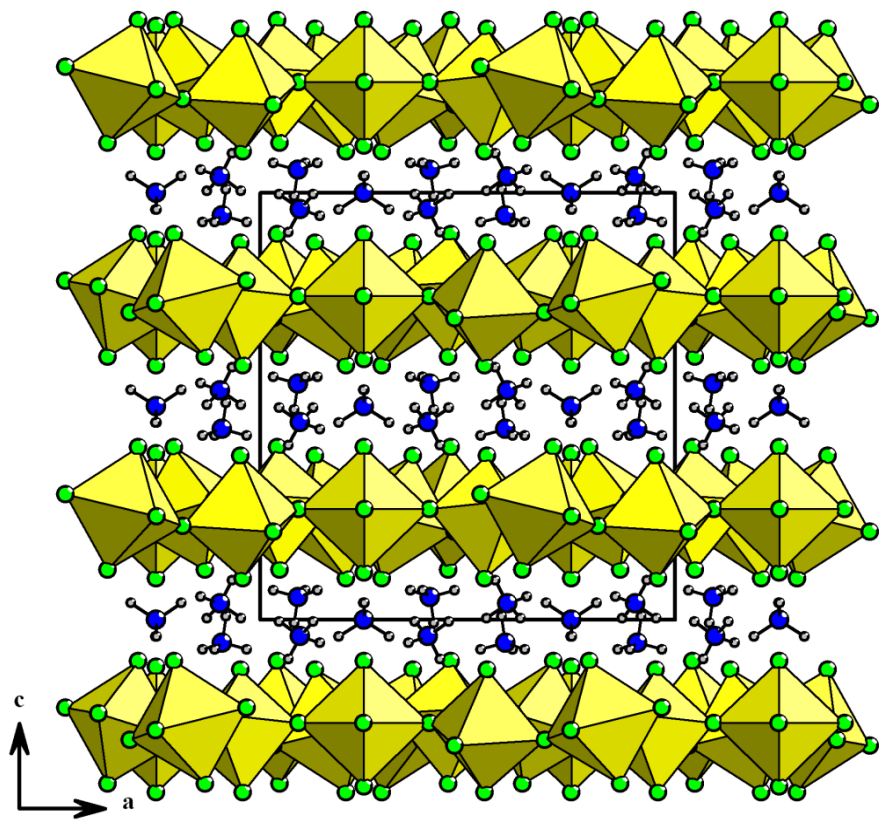

**Figure S7.** Structure of **2**, showing the TTB-type layers separated by ammonium cations.

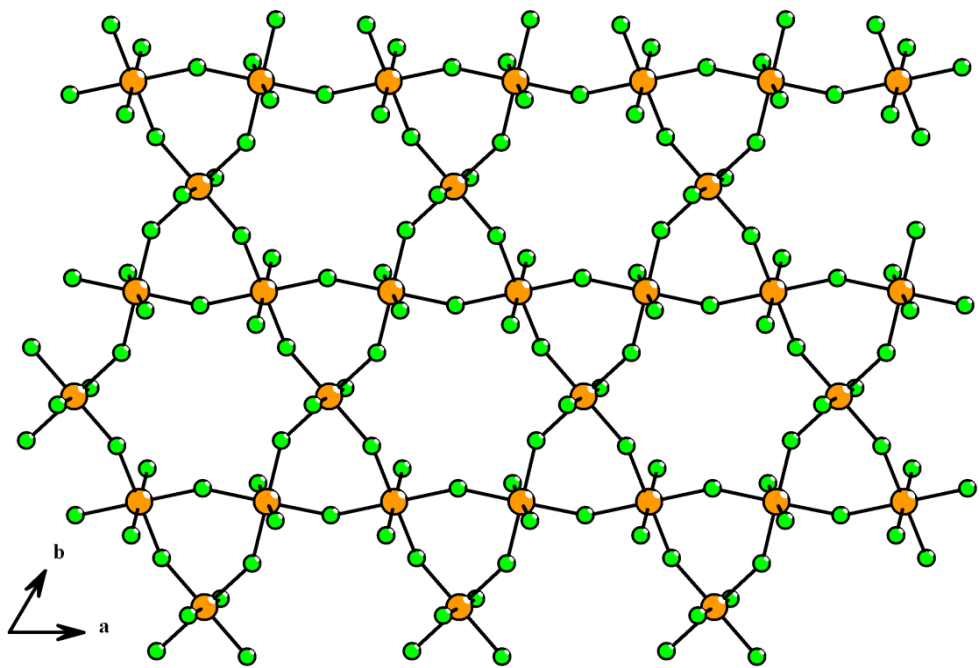

**Figure S8.** Kagome layer found in **3**.

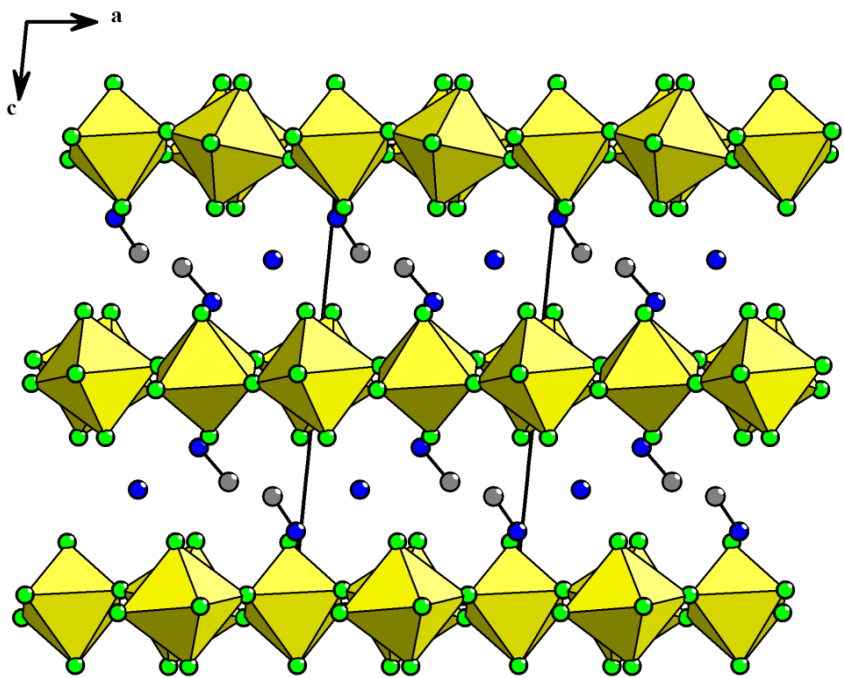

**Figure S9.** Structure of **3**, showing the kagome-type layers separated by methyl ammonium and ammonium cations.

Colour scheme used in all pictures: Vanadium (Orange), Fluorine (green), Nitrogen (dark blue), Carbon (dark grey), Hydrogen (light grey).